\documentclass{aa}
\usepackage{natbib}
\usepackage{graphicx}
\usepackage{txfonts}
\newcommand{\ks}{KS\,1731--260}

\begin{document}
   \title{Optical/infrared observations of the X-ray burster KS1731--260 in quiescence}


   \author{C. Zurita
          \inst{1,2}
          \and
          E. Kuulkers\inst{3}
           \and
          R.M. Bandyopadhyay\inst{4}
           \and
          E.M. Cackett\inst{5}\fnmsep\thanks{Chandra Fellow}
           \and
          P.J. Groot\inst{6}
           \and
          J.A. Orosz\inst{7}
           \and
          M.A.P. Torres\inst{8}
           \and
          R. Wijnands\inst{9}
          }

   \institute{Instituto de Astrof\'{\i}sica de Canarias,
              C/V\'{\i}a L\'actea s/n; 3800 La Laguna, Spain
              \email{czurita@iac.es}
         \and
             Departamento de Astrof\'{\i}sica, Universidad de La Laguna,
             38205 La Laguna, Tenerife, Spain
         \and
             ESA/ESAC, Villafranca del Castillo, P.O. Box 50727, 28080
             Madrid, Spain
         \and
         Department of Astronomy, University of Florida, Gainesville, 
         FL 32611, USA
         \and
         Department of Astronomy, University of Michigan, 500 Church Street,
         Ann Arbor, MI 48109-1042, USA
         \and
         Department of Astrophysics, Radboud University Nijmegen, PO Box
         9010, 6500 GL Nijmegen, the Netherlands 
         \and
         Department of Astronomy, San Diego State University, 5500 Campanile
         Drive, San Diego, CA 92182, USA
         \and
         Harvard-Smithsonian Center for Astrophysics, 60 Garden Street,
         Cambridge, MA 0213, USA
         \and
         Astronomical Institute "Anton Pannekoek," University of Amsterdam,
         Kruislaan 403, 1098 SJ Amsterdam, The Netherlands
             }
   \date{Received September 15, 1996; accepted March 16, 1997}

 
\abstract
  {}
  {We performed an  optical/infrared study of the counterpart  of  the
  low-mass X-ray binary KS\,1731--260  to test its  identification and obtain 
  information about the donor.}
  {Optical and infrared images of  the counterpart of KS\,1731--260 were taken
  in  two different  epochs  (2001 and  2007)  after the  source returned  to
  quiescence in X-rays. We compared those observations with obtained when
  KS\,1731--260 was still active.}
  {We   confirm  the   identification  of
  KS\,1731--260  with  the previously  proposed  counterpart  and improve  its
  position to  $\alpha$=17:34:13.46 and $\delta$=-26:05:18.60.   The $H$-band
  magnitude  of  this candidate  showed  a  decline  of $\sim$1.7\,mags  from
  outburst  to quiescence.  In  2007 April  we obtained  $R$=22.8$\pm$0.1 and
  $I$=20.9$\pm$0.1 for KS\,1731--260.   Similar optical brightness was measured
  in June 2001 and July 2007.   The intrinsic optical color $R-I$ is consistent
  with spectral types from F to G for the secondary although there is a large
  excess over that  from the secondary at the  infrared wavelengths. This may
  be due to emission from the cooler outer regions of the accretion disk.  We
  cannot rule  out a brown dwarf as  a donor star, although  it would require
  that  the distance to  the source  is significantly  lower than  the 7\,kpc
  reported by Muno et al. 2000.}
{}



   \maketitle
%

\section{Introduction}

Low-mass X-ray  binaries are  systems  in which  a low-mass companion  star
transfers  material onto  a neutron  star or  black hole.   In  these compact
binaries, the intense X-ray irradiation usually overwhelms the light from the
donor  \citep[e.g.][]{charles06}.  There  are,  however,  some  systems,  the
so-called   X-ray   transients,   in   which   substantial   X-ray   activity
(10$^{36}$-10$^{38}$erg\,s$^{-1}$) only  occurs  during   well-defined
outbursts. The outbursts typically last  from weeks to months and are usually
separated by long  intervals (years to decades) of  very low X-ray luminosity
(10$^{30}$-10$^{34}$erg\,s$^{-1}$). During these  intervals of quiescence, the
emission from the  accretion flow fades to the point the companion star
is clearly visible and is nearly undisturbed by irradiation; hence, it can
be studied to derive the parameters of the binary. Most of the low-mass X-ray
binaries have  orbital periods of  a few hours  to days and  contain ordinary
hydrogen-rich  donor stars.  The so-called  ultra compact  binaries, however,
have orbital periods shorter than 80 minutes. The small period implies such a
small   Roche   lobe   that   the   donor  star   must   be   hydrogen   poor
\citep[e.g.][]{nelson86}.


The transient  \ks\ was  discovered with the  Mir/Kvant instrument  in August
1989 \citep{sunyaev89}.  The presence of  type-I X-ray bursts coming from the
system indicates that its compact  object is a neutron star \citep{sunyaev89,
  sunyaev90}  and places  an upper  limit on  the distance  to the  source of
7\,kpc \citep{muno00} and  7.8\,kpc \citep{galloway08} assuming  a pure
  helium photosphere  for a 1.4\,M$_{\odot}$  and R=10\,Km neutron  star. The
  corresponding  distances for  a 2\,M$_{\odot}$  neutron star  would  be 9\%
  greater. The \citet{galloway08} best estimations are 7.2$\pm$1\,kpc for pure
  helium and  5.6$\pm$0.7\,kpc for material with  cosmic abundances (hydrogen
  fraction  X=0.7).  In  contrast to  most  X-ray transients,  \ks\ did  not
disappear after a few weeks to  months, but it could be observed continuously
at  high luminosities.   However,  in February  2001,  after having  actively
accreted for over a decade, the  source suddenly turned off.  A {\it Chandra}
observation of the source was performed  a few months after this event and an
X-ray  luminosity  of  only  $2\times10^{33}erg\,s^{-1}$  could  be  measured
\citep{wijnands01a}.

Several  authors  tried   to  identify   the  optical
counterpart  of \ks\ during  its long  active episode.   In the  error circle
obtained    with   Kvant,    many   optical    stars   were    present,   but
\citet{cherepashchuk94}  identified two  promising stars  on the  red Palomar
Survey plate.  However, using  the significantly higher spatial resolution of
the {\it ROSAT/HRI}, \citet{barret98} demonstrated that those stars could not
be identified with \ks\ and suggest 13 possible candidates, which were later
ruled   out  by   the  analysis   of  the   {\it  Chandra}   image   made  by
\citet{revnivtsev02}.  These  authors propose a  likely counterpart, although
it  could not  be  conclusively identified because of the lack  of 
observations  in quiescence.  Finally,  \citet{wijnands01b} made o
bservations  when the source
turned off  and identified  the counterpart  of \ks\ as  a very  weak optical
source in  the {\it  Chandra} error circle.   However, no  optical magnitudes
could be measured.   The only optical/near-IR magnitude measured  for \ks\ in
quiescence was obtained by \citet{orosz01} who reported $J$=18.62$\pm$0.21 in
2001 July 13.

In  this  paper we  present  the optical  and  infrared  observations of  the
counterpart of \ks. This is an  important issue since this source is only one
of  a few  sources  where crustal  cooling  has been  observed in  quiescence
\citep[][and  references therein]{cackett06}.  Knowledge  of the  neutron star
mass helps to set the timescale of this cooling, as higher mass neutron stars
have  a thinner  crust, hence would  cool more  quickly \citep{brown09}.  In
addition,   \ks\   is   one   of   the   few   sources   showing   superburts
\citep{kuulkers02}.  Therefore, any information  on the  nature of  the donor
star  may give  us a  clue to  what kind  of material  is accreted  onto the
neutron star.



\section{Observations and reductions}

Infrared  $H$ and  $K$ images  of  \ks\ were  obtained on  the 3.8\,m  United
Kingdom Infrared Telescope (UKIRT) with the UKIRT Fast Track Imager (UFTI) on
UT  2001  July  9.   Seeing  during  the observations  was  measured  at  0.9
arcsec. For both the $H$- and $K$-band images of the target, a series of five
consecutive 60\,s exposures were obtained with offsets of 20\,arcsecs between
each exposure.  Observations of a  photometric standard star were obtained in
a similar way, with 5\,s exposure  time and using a subarray of the detector
(47\,arcsec$^2$ field  of view). For both  the target and  the standard stars,
the position of  the object on the array was moved  between exposures so that
the group could be median-stacked to produce a sky flat. Basic data reduction
was  performed  using the  ORAC-DR  online  reduction  system at  UKIRT.  The
magnitudes were derived from IRAF/daophot point-spread function (psf) fitting
with an aperture correction.

Optical images of  the field of \ks\ in Sloan $r'$,  $i'$ and $z'$ \citep[see
  e.g.][]{fukugita96} were obtained on UT  2001 June 28 on the 6.5\,m Magellan
Walter  Baade telescope  at Las  Campanas Observatory  (Chile) and  the MagIC
camera with  600\,s of  exposure time  in each filter.   We also  observed in
Johnson/Bessel $R$ and $I$  on UT 2007 April 26 and 27  with the same telescope
but equiped  with the IMACS  detector and with  900\,s of total  exposure in
each band. Finally, we obtained images  in Johnson/Bessel $R$ and Gunn $i$ on
UT 2007 June 16 on the 3.6\,m  telescope at La Silla Observatory in Chile with
1800\,s exposure time in each filter.\\

In none of  the above nights were standard stars observed,  so we calibrated a
set of 15 faint  stars in the field of view of \ks\  in an independent run on
UT 2007 April 25  on the 1.5\,m telescope at San  Pedro M\'artir Observatory in
Mexico.   We performed  a  color-dependent $BVRI$  calibration using  several
standard stars from four  Landolt plates \citep{landolt92}. The conversion to
the Sloan  filter set was made  using the empirical  transformations given in
\citet{jordi06}.  \ks\ was not detected that night. In the rest of the nights
we performed aperture  photometry on our object and in  the set of previously
calibrated comparison stars.  All the  optical images were corrected for bias
and flat-fielded in the standard way using IRAF
\footnote   {IRAF  is   distributed   by  the   National  Optical   Astronomy
Observatories,  which are  operated by  the Association  of  Universities for
Research in  Astronomy, Inc., under  cooperative agreement with  the National
Science   Foundation.}    tasks.   An   observing   log   is   presented   in
Table~\ref{table:log}.

\begin{table}
\caption{Log of the observations}         
\label{table:log}      
\centering                          
\begin{tabular}{c c c c}        
\hline\hline                 
Date & n$\times$exp.time (s) & Filter & Telescope \\    
\hline                        
2001 June 28 & 1$\times$600 & $r'$     & 6.5\,m Magellan  \\
             & 1$\times$600 & $i'$    &  \\
             & 1$\times$600 & $z'$    &  \\
2001 July 09  & 5$\times$60 & $H$     & 3.8\,m UKIRT\\
             & 5$\times$60 & $K$     &  \\
2007 April 25  & 3$\times$900 & $V$     & 1.5\,m San Pedro M\'artir \\
             & 3$\times$600 & $R$     &  \\
             & 3$\times$600 & $I$     &  \\ 
2007 April 26  & 3$\times$300 & $R$     &  6.5\,m Magellan \\
2007 April 27  & 3$\times$300 & $I$     &  6.5\,m Magellan \\
2007 July 16  & 1$\times$1800 & $R$     & 3.6\,m La Silla \\
             & 1$\times$1800 & $i$     &  \\

\hline                                   
\end{tabular}
\end{table}

\section{The position of \ks}

The {\it  Chandra}/ACIS-S observation on \ks\  was obtained on  2001 March 27
00:17-06:23 UTC (only  a few months after the source turned  off) for a total
onsource time of  $\sim$20\,ksec. For details about this  observation and the
discussion of the result obtained from  the spectral analysis of the data, we
refer  the reader  to \citet{wijnands01a,  wijnands01b}. Two  CIAO  tools are
usually  used to determine  the presence  of X-ray  sources in  {\it Chandra}
fields and to  obtain their positions: {\it celldetect}  and {\it wavdetect}.
Both   tools   detected  two   X-ray   sources   in   our  field   \citep[see
  also][]{wijnands01a,  wijnands01b}, but  although the  obtained coordinates
were very similar  for both tools, they differ  slightly (by 0.1\,arcsec). We
also used  the IRAF tool {\it  daofind} to obtain the  coordinates, and these
coordinates were slightly different from  those obtained with the CIAO tools.
The statistical errors  on the position are very small, but  the spread in the
coordinates  as obtained  with the  different  tools can  be used  as a  good
indication of  the accuracy of the  coordinates. As expected,  the spread is
greater for the extra {\it Chandra} source (designated CXOU\,J173412.7-260548)
because of its lower number of counts compared to \ks.

The pointing accuracy  of the satellite is approximately  0.6 arcsec, but the
astrometric accuracy of the coordinates can  be improved if it is possible to
tie the  {\it Chandra} coordinate  frame with others  (such as 2MASS  with an
astrometric    accuracy     0$.\!\!^{\prime\prime}$2).     We    only    have
CXOU\,J173412.7-260548 to  try to obtain better astrometric  accuracy, but in
the  2MASS catalog  a star  (2MASSI J173412.7-260548)  was present  with very
similar coordinates to CXOU\,J173412.7-260548;  the coordinates are 0.4 arsec
off.  Therefore, we identify CXOU\,J173412.7-260548 with the 2MASS star. This
is  the only  star above  the 2MASS  detection limit  into the  Chandra error
circle. To estimate the  probability that an  unrelated source has
fallen in the X-ray error circle by chance, we note that there are 7 stars of
similar brightness or brighter to 2MASSI J173412.7-260548 within a 1 arcmin
circle,  so we  estimate a  7$\times$10$^{-4}$ probability  that  an unrelated
object  is  falling  by  chance  into the  0.6\,arcsec  Chandra  error  circle.
CXOU\,J173412.7-260548 has an offset  with respect to 2MASSI J173412.7-260548
of 0.016, 0.024, and 0.001 sec in  right ascension and 0.12, 0.03, and 0.15 sec
in declination  for the {\it  celldetect}, {\it wavdetect}, and  {\it daofind}
tool, respectively.  We applied the same offsets for the position obtained for
\ks. By combining  these  offsets  we  derived a  best  position  of  \ks\  of
$\alpha$=17:34:13.47 and $\delta$=-26:05:18.8, with an accuracy of 0.4\,sec.

\begin{figure}
  \centering
   \includegraphics[width=14cm]{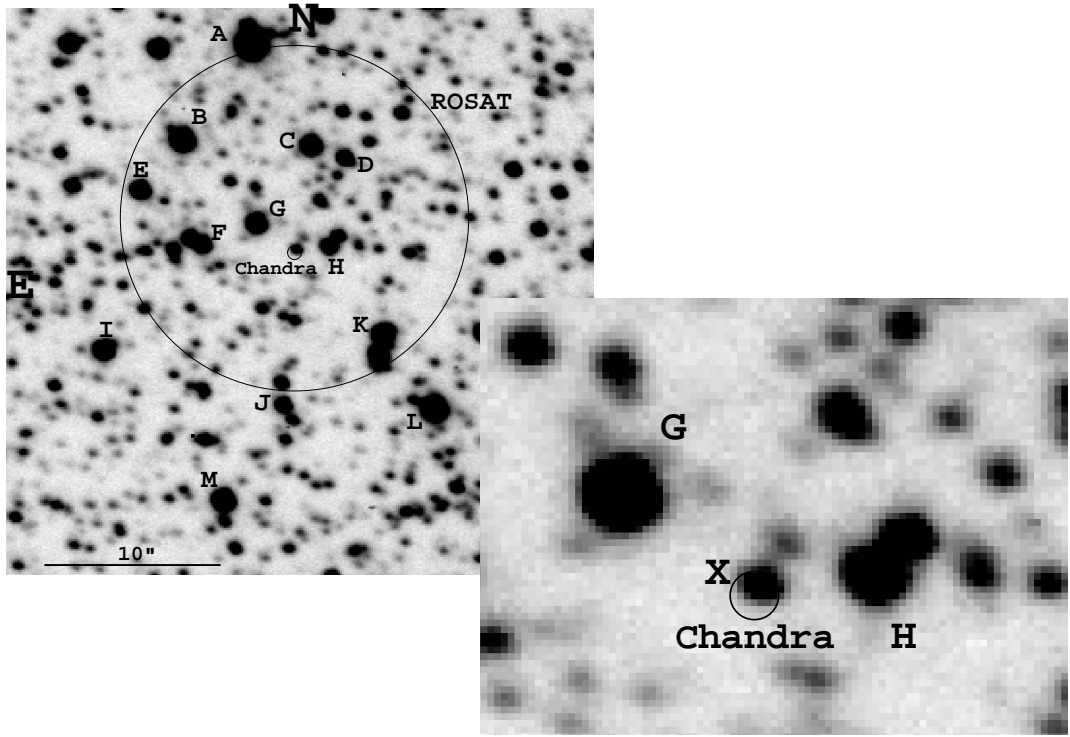}
     \caption{A Magellan  $I$ image  of  the ROSAT  error box  (10\,arcsec
radius) of \ks\  (left) and a close-up of  the Chandra position (0.4\,arcsec)
with star `X' as our  proposed optical and infrared counterpart (right). Star
labeling is as in Barret et al. (1998).  North is upwards and east is to 
the left.}
         \label{figure:field}
   \end{figure}

In Fig.\ref{figure:field} we  show the Magellan $I$ image,  taken  2007 April
25, with  the ROSAT and {\it  Chandra} 0.4\,sec diameter  error circle.  To
obtain  a  precise  astrometric  solution,  we  used  the  positions  of  the
astrometric   standards  selected  from   the  USNO-B1   astrometric  catalog
\footnote  {USNO-B1  is currently  incorporated  into  the Naval  Observatory
  Merged  Astrometric   Data-set  (NOMAD)  which   combines  astrometric  and
  photometric  information  from   Hipparcos,  Tycho-2,  UCAC,  Yellow-Blue6,
  USNO-B,  and  the  2MASS,  www.nofs.navy.mil/data/fchpix/} with  a  nominal
0$.\!\!^{\prime\prime}$2  uncertainty.  A  hundred reference  objects  can be
identified in  our field but, to minimize  potential positional uncertainties
caused by overlapping  stellar profiles, we selected only  28 isolated stars,
discarding  the  stars  with  significant  proper motions.   The  IRAF  tasks
ccmap/cctran were  applied for the astrometric transformation  of the images.
Formal  rms  uncertainties  of  the   astrometric  fit  for  our  images  are
$\la$0$.\!\!^{\prime\prime}$15 in both right ascension and declination, which
is compatible with  the maximum catalog position uncertainty  of the selected
standards.  It  is clear that only  one viable candidate  counterpart is left
within  the {\it  Chandra} error  box.  This  image, taken  with  good seeing
conditions ($\sim$ 0$.\!\!^{\prime\prime}$5), reveals  a faint nearby star at
northwest of the  target.  Finally, we measured the center  of the source 'X'
by fitting a Gaussian to the star profile and improve the position of \ks\ to
$\alpha$=17:34:13.46 and  $\delta$=-26:05:18.60 with a  conservative estimate
of our  3$\sigma $ astrometric uncertainty  of $\la$0 $.\!\!^{\prime\prime}$2
in both RA and Dec.

\section{Photometry}

\begin{figure}
  \centering
   \includegraphics[width=14cm]{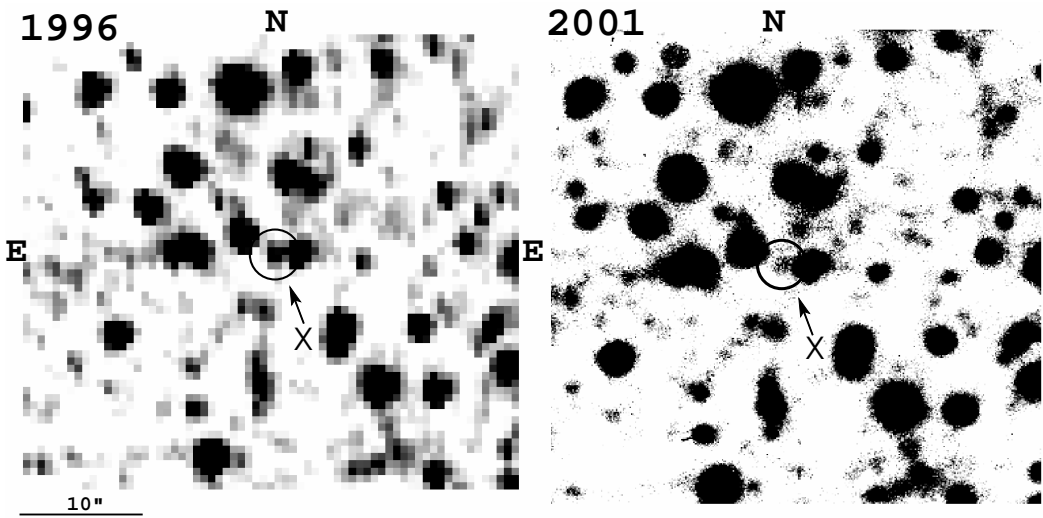}
     \caption{$H$-band images  of the  field around the  proposed counterpart
       'X'  obtained   in  1996  with   CFHT  (left)  and  2001   with  UKIRT
       (right). North is upwards and East is to the left.}
         \label{figure:compH}.
   \end{figure}

Infrared $H$  and $K$ images of \ks\  in quiescence were obtained  on UT 2001
July 9. We compared the brightness of the stars in the field of \ks\ obtained
on  1996 June  by  \citet{barret98}  with our  2001  UKIRT observations  (see
Fig.\,\ref{figure:compH}).   In  Table  \ref{table:Hmags}  we  give  the  old
$H$-band magnitudes, from the  original \citet{barret98} paper, and new ones,
as  well  as  our  $(H-K)$  2001  colors.  Stars  have  been  labeled  as  in
\citet{barret98}.   We   also  included   the  most  likely   counterpart  of
\ks\ (source `X').   The source `X' was not  measured by \citet{barret98}, so
we determined its $H$-band  magnitude from the original observations. Because
of the faintness of target `X',  we first cleaned the contamination of nearby
stars by using the IRAF allstar task,  which subtracts the best PSF fit of the
set of contaminating stars.  This step was iterated to minimize the residuals
after PSF  subtraction.  Although there is a  scatter of 0.17 rms  in the $H$
band magnitudes as  derived from the \citet{barret98} CFHT  image and the new
UKIRT observations,  only target `X', of  the sources near  the Chandra error
circle, shows a substantial variation.  For this object $H$=17.70$\pm$0.20 in
2001, that is,  it has declined by $\sim$1.7\,mag  between both observations.
The large change in $H$-band magnitude between burst and quiescence, together
with its location in the {\it Chandra} error circle, leads us to conclude that
source `X' is indeed the infrared counterpart to \ks.

\begin{table}
\caption{$H$-band  magnitudes and  colors of  isolated stars  near  the {\it
    Chandra} error circle.  CFHT and UKIRT observations were obtained on 1996
  June  and  2001  July,  respectively.  Star  labeling is  an  in  Barret  et
  al. (1998) and Fig.~\ref{figure:field}}
\label{table:Hmags}      
\centering                          
\begin{tabular}{c c c c}        
\hline\hline                 
Star & $H$(CFHT) & $H$(UKIRT) & $(H-K)$(UKIRT) \\    
\hline                        
A &   12.35$\pm$0.06  &  11.91$\pm$0.10  & 0.71$\pm$0.10\\
B &   13.10$\pm$0.08  &  13.12$\pm$0.10  & 0.70$\pm$0.10\\
C &   14.14$\pm$0.19  &  14.01$\pm$0.10  & 0.71$\pm$0.10\\
D &   14.79$\pm$0.92  &  15.29$\pm$0.15  & 0.71$\pm$0.15\\
E &   14.44$\pm$0.15  &  14.33$\pm$0.10  & 0.76$\pm$0.10\\
F &   14.00$\pm$0.12  &  14.33$\pm$0.11  & 0.76$\pm$0.11\\
G &   13.84$\pm$0.12  &  13.89$\pm$0.10  & 0.76$\pm$0.10\\
H &   14.99$\pm$0.34  &  15.07$\pm$0.11  & 0.72$\pm$0.11\\
I &   14.38$\pm$0.12  &  14.24$\pm$0.10  & 0.70$\pm$0.10\\
J &   15.14$\pm$0.54  &  15.25$\pm$0.15  & 0.75$\pm$0.15\\
K &   13.64$\pm$0.09  &  13.83$\pm$0.10  & 0.77$\pm$0.10\\
L &   13.19$\pm$0.08  &  12.94$\pm$0.10  & 0.67$\pm$0.10\\
M &   13.39$\pm$0.06  &  13.18$\pm$0.12  & 0.79$\pm$0.12\\
{\bf X} & {\bf16.00}$\pm${\bf0.60} & {\bf17.70}$\pm${\bf0.20} & {\bf0.70}$\pm${\bf0.30}\\
\hline                                   
\end{tabular}
\end{table}


We observed  \ks\ in  three  different epochs  when the  target was  in
quiescence: in 2001 in Sloan $r'$,$i'$,$z'$,  in April 2007 in Bessel $R$ and
$I$,  and in  June  2007  in Bessel  $R$  and Gunn  $i$.   To obtain  reliable
magnitude measurements of our very faint target, we cleaned the contamination
of the stars near \ks\  (G and H in Fig.~\ref{figure:field}) by subtracting
the best PSF. Photometric error estimates on the optical magnitudes are based
on  a combination of  Poisson statistics  and the  error contribution  of the
stars used for  calibration.  For the optical magnitudes  in different
nights, we refer to Table~\ref{table:mags}.

\section{Discussion}

Table~\ref{table:mags}  summarizes   the  magnitudes  of   \ks\  measured  in
quiescence  including the  $J$  magnitude of  \citet{orosz01}.   The $J$  and
$H$-band  magnitudes of  this candidate  show  a decline  of $\sim$1.3  and
1.7\,mags, respectively, from outburst (1996 June) to quiescence (2001 July).  In
the $K$-band,  however, the source  is only $\sim$ 0.6\,mag  fainter compared
with  the $K'$  magnitude obtained  by \citet{mignani02}  in 1998  July.  This
could mean either  that the system faded in infrared from  1996 to 1998, when
it was  actively accreting or that the  disk emission  is lower in  $K$ so,
after the X-rays turn off, the disk  has become much less luminous,  so the
$J$ and $H$ magnitudes dropped.

\begin{figure}
  \centering
   \includegraphics[width=14cm]{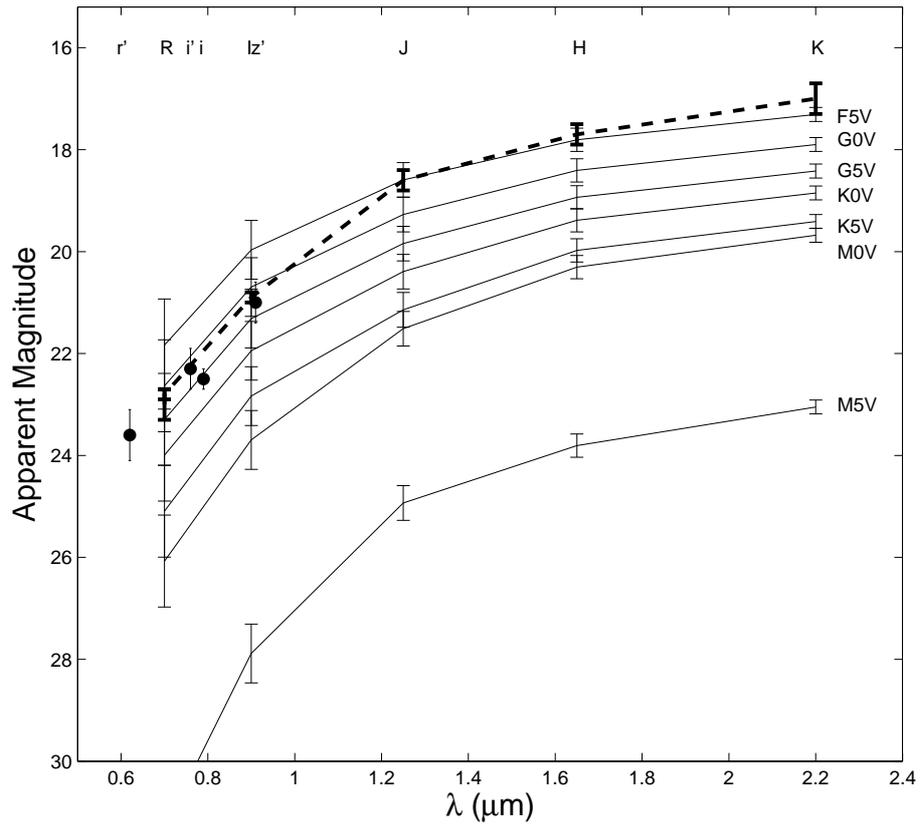}
     \caption{$RIJHK$ SED for various type stars and \ks\ (dashed line).}
         \label{figure:sed}
   \end{figure}

In 2007 \ks\ had the same  optical brightness, within the errors, as in 2001.
To  determine  the nature  of  its companion,  we  plot the  $RIJHK$
spectral energy distribution (SED) for various main-sequence type stars along
with that of \ks\ (Fig.~\ref{figure:sed}).  The absolute magnitudes and the
colors  of the stars  are taken  from \citet{leggett92}  and \citet{allen76},
whereas the apparent magnitudes were  calculated using the extinction laws of
\citet{cardelli89}  assuming  a reddening  of  A$_V\sim$6  obtained from  the
spectral    fits   to    the   combined    {\it    Chandra/XMM-Newton}   data
\citep{wijnands02}   and   a    distance   of   7\,kpc   \citep[upper   limit
  from][]{muno00}. The  error bars for these apparent  magnitudes account for
the  uncertainty  in the  reddening.  The optical  color  $R-I$  of \ks\  is
consistent with  spectral types  from F to  G. However, the  infrared colors
($J-K$=0.9 and  $H-K$=0.6) are much  higher than expected  even for a  late M
star.

Observations of transients in  quiescence have predominately been carried out
in the optical,  and in this wavelength range the accretion  disk is known to
contribute significantly to the observed  flux.  Typically it is assumed that
the  disk  spectrum  is  a  featureless  continuum  and  that  it  marginally
contributes to  the overall light  in the infrared.  Therefore,  many authors
have  used  infrared observations,  rather  than  optical,  to determine  the
ellipsoidal  variability and  constrain the  mass ratio  and  the inclination
angle  in these  systems  \citep[see e.g.][for  a  review]{charles06}. If  we
consider the infrared data  alone, assuming that the near-infrared magnitudes
are completely dominated by the light  of the secondary, we cannot rule out a
brown dwarf as donor star \citep[$J-K\ge$1;][]{cruz09}.  The possibility that
\ks\  is  an  extremely  narrow  binary  system with  an  orbital  period  of
$\sim$1\,h has  been suggested by  \citet{muno00} based on their  analysis of
the burst  pulsations of this  system and the  spin-frequency interpretation.
\citet{kuulkers02} also  report a  twelve-hour long  X-ray flare  from this
source.  \citet[][see  also \citet{strohmayer02}]{cumming01} pointed  out the
unstable carbon burning in a  layer beneath the (un)stable hydrogen/helium or
helium layer, deeper in the neutron  star, as a possible mechanism to explain
these events. Obviously,  for  this to work, the ashes of the burning
hydrogen/helium layer need  to contain carbon. One way to  achieve this is to
have stable  burning of helium   so the donor star  should be helium-rich.
However, a  brown dwarf donor would  require that the distance  is lower than
500\,pc,  well below the  estimation based  on radius-expansion  X-ray bursts
\citep{muno00, galloway08}.

\begin{figure}
  \centering
   \includegraphics[width=10cm]{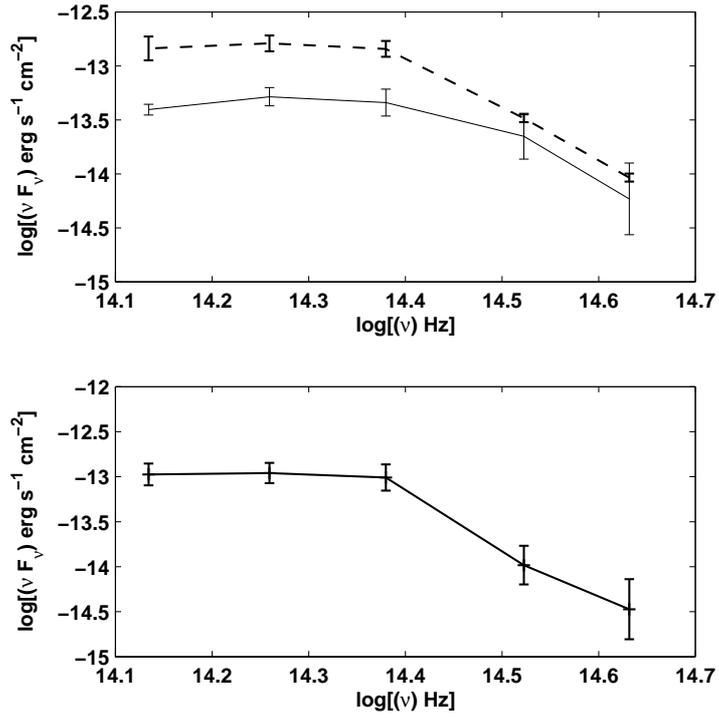}
     \caption{Top panel: SEDs of \ks\ (dashed line) and a G5V star (solid
       line) assuming  a reddening  of A$_V\sim$6 and  a distance  of 7\,kpc.
       Bottom  panel:  spectrum  of  the nonstellar  component  obtained  by
       subtracting the G5V colors from the \ks\ colors.}
         \label{figure:subsed}.
   \end{figure}

Nevertheless, there is  no reason to argue that  the nonstellar contribution
to the  near infrared flux  is minimal and consistent  throughout quiescence.
On the contrary,  optically thick material (hence  thermal infrared flux)
at  the  outer regions  of  the  accretion  disk is  theoretically  plausible
\citep{hynes05}.  Furthermore, infrared flickering has been recently detected
in   the   X-ray   Transients    (and   black   hole   candidates)   J0422+32
\citep{reynolds07} and A0620-00 \citep{cantrell08} in quiescence. In A0620-00
system a  jet was  seen in the  radio band \citep{gallo06}.   The synchrotron
spectrum, which  is thought to  be the jet  signature, is frequently  seen at
radio  but  also up  to  higher frequencies  in  the  infrared. Finally,  the
analysis  of the  SEDs  of a  sample  of quiescent  transients  (one of  them
containing a neutron star) have shown  an infrared excess probably due to the
presence of  a cool disk  component \citep{reynolds08}. Hence, a  nonstellar
infrared  component superimposed on  a main  sequence, F  to G  type, stellar
spectrum  can explain  our SED.  The  distance to  the system  would be  then
$\sim$5\,kpc for  a late G  star and $\sim$12\,kpc  for a late F.  Assuming a
distance to the source of 7.2\,kpc \citep{galloway08} the optical counterpart
is more consistently a early G  type star.  To recover the broadband spectrum
of  the nonstellar  component we  subtracted the  companion star,  using the
apparent magnitudes  of a G5V star  assuming a reddening of  A$_V\sim$6 and a
distance  of 7\,kpc, from  the \ks\  colors (figure  \ref{figure:subsed}, top
panel).  Again, the error bars  for these apparent magnitudes account for the
uncertainty   in   the    reddening.    The   resultant   component   (figure
\ref{figure:subsed},    bottom   panel)   is    far   from    the   canonical
F$_{\nu}\propto\nu^{1/3}$  accretion  disk  spectrum.  However,  fitting  the
infrared  alone we  find  a flat  spectrum  with $\alpha$=-0.1$\pm$0.1  where
F$_{\nu}$=$\nu^{\alpha}$.   The very  flat infrared  SED (F$_{\nu}\sim$const)
could  naturally be  interpreted  as a  mixture  of an  optically thick  disk
spectrum and  flat spectrum emission,  possibly synchrotron \citep{fender00}.
Therefore, we favor an H-rich system instead of an ultra compact binary.

Combining \citet{paczynski71}  expression for the averaged radius  of a Roche
lobe with Kepler's  Third Law, we get the  well-known relationship between the
secondary's      mean      density      and     the      orbital      period:
$\rho=(110/P_h^2)$\,g\,cm$^{-3}$, where $\rho$ is  the mean density and $P_h$
the orbital  period in  hours. From this  expression we estimated  an orbital
period of about 10\,hr for a G type star.

\begin{table}
\caption{Magnitudes of \ks\ for the three epochs in quiescence.}        
\label{table:mags}      
\centering                          
\begin{tabular}{c l }        
\hline\hline                 
Date & mag \\    
\hline                        
2001 June 28 &   $r'$=23.6$\pm$0.4 \\
         &      $i'$=22.3$\pm$0.4 \\
         &      $z'$=21.0$\pm$0.4 \\
2001 July 09 &   $H$=17.7$\pm$0.2 \\
         &      $K$=16.7$\pm$0.3 \\
2001 July 13 &   $J$=18.6$\pm$0.2 (*)\\
2007 April 26 &   $R$=22.8$\pm$0.1 \\
2007 April 27 &   $I$=20.9$\pm$0.1 \\
2007 July 16&    $R$=23.0$\pm$0.3 \\
         &      $i$=22.5$\pm$0.2 \\
\hline
 {\footnotesize (*) magnitude $J$ from Orosz et al. 2001} & \\
\end{tabular}
\end{table}                                   

\onecolumn

\twocolumn

\begin{acknowledgements}
We thank Christian Motch for providing the archival CFHT data. CZ is grateful
to  Celia  S\'anchez and  for  the hospitality  of  the  ESA, European  Space
Astronomy  Centre (ESAC).   EMC gratefully  acknowledges support  provided by
NASA through the Chandra Fellowship Program, grant number PF8-90052.
\end{acknowledgements}

\bibliographystyle{aa}
\bibliography{czurita}

\end{document}